\documentclass[sigplan,nonacm]{acmart}

\acmJournal{PACMPL}
\acmVolume{1}
\acmNumber{CONF} %
\acmArticle{1}
\acmYear{2018}
\acmMonth{1}
\acmDOI{} %
\startPage{1}

\setcopyright{none}

\usepackage{booktabs}   %
\usepackage{subcaption} %

\usepackage[utf8]{inputenc}
\usepackage[english]{babel}
\usepackage{xparse}
\usepackage{xspace}
\usepackage{thmtools}
\usepackage[final]{fixme} %
\usepackage{xcolor}
\usepackage{mathpartir}
\usepackage[]{microtype}
\usepackage{multirow}
\usepackage{makecell}
\usepackage{marvosym}
\usepackage{wasysym}
\usepackage{pifont}
\usepackage{mathtools}
\usepackage{stmaryrd}
\usepackage{scalerel}
\usepackage{tensor}
\usepackage{xifthen}
\usepackage{iris}
\usepackage{heaplang}
\usepackage{pftools}
\usepackage{soul}
\usepackage{tikz}
\usepackage{minted}
\usepackage{threeparttable}
\usepackage{fontawesome}
\usepackage{enumitem}
\usetikzlibrary{calc, shapes, arrows, automata, patterns}

\makeatletter
\addto\extrasenglish{%
  \renewcommand*\chapterautorefname{\S\@gobble}
  \renewcommand*\sectionautorefname{\S\@gobble}
  \renewcommand*\subsectionautorefname{\S\@gobble}
  }
\makeatother
\newcommand*{\myeqref}[2][]{%
  \hyperref[{#2}]{#1(\ref*{#2})}%
}

\renewcommand*{\lineref}[1]{\hyperref[#1]{line~\ref*{#1}}}
\newcommand*{\linerangeref}[2]{\hyperref[#1]{lines~\ref*{#1}}\hyperref[#2]{-\ref*{#2}}}
\newcommand{\mypageref}[1]{\hyperref[#1]{page~\pageref*{#1}}}

\newboolean{appendixincluded}
\setboolean{appendixincluded}{false}

\newcommand{\appendixdocname}{\ifthenelse{\boolean{appendixincluded}}{appendix}{companion appendix} }
\newcommand{\appendixsect}[2]{\ifthenelse{\boolean{appendixincluded}}{(\autoref{#1})}{\cite[Section #2]{Artifact}}}
\newcommand{\appendixref}[2]{\ifthenelse{\boolean{appendixincluded}}{\autoref{#1}}{the companion appendix \cite[Section #2]{Artifact}}}

\setminted{escapeinside=\^\^, mathescape=true, linenos=true, numbersep=5pt, framesep=2mm, fontsize=\footnotesize, extrakeywords={effect, return, fix_effect, partial, coinductive_fixpoint, monotonicity}}

\DeclareUnicodeCharacter{2264}{\ensuremath{\le}}
\DeclareUnicodeCharacter{2208}{\ensuremath{\in}}
\DeclareUnicodeCharacter{2203}{\ensuremath{\exists}}
\DeclareUnicodeCharacter{25C1}{\ensuremath{\triangleleft}}
\DeclareUnicodeCharacter{2097}{\ensuremath{{}_l}}
\DeclareUnicodeCharacter{2260}{\ensuremath{\neq}}
\DeclareUnicodeCharacter{2205}{\ensuremath{\emptyset}}
\DeclareUnicodeCharacter{222A}{\ensuremath{\cup}}
\DeclareUnicodeCharacter{228E}{\ensuremath{\uplus}}
\DeclareUnicodeCharacter{2200}{\ensuremath{\forall}}
\DeclareUnicodeCharacter{25A1}{\ensuremath{\always}}
\DeclareUnicodeCharacter{03B3}{\ensuremath{\gamma}}
\DeclareUnicodeCharacter{03C1}{\ensuremath{\rho}}
\DeclareUnicodeCharacter{03BB}{\ensuremath{\lambda}}
\DeclareUnicodeCharacter{03B1}{\ensuremath{\alpha}}
\DeclareUnicodeCharacter{03B2}{\ensuremath{\beta}}
\DeclareUnicodeCharacter{03B9}{\ensuremath{\iota}}
\DeclareUnicodeCharacter{2295}{\ensuremath{\oplus}}
\DeclareUnicodeCharacter{03C3}{\ensuremath{\sigma}}
\DeclareUnicodeCharacter{2091}{\ensuremath{{}_e}}
\DeclareUnicodeCharacter{2081}{\ensuremath{{}_1}}
\DeclareUnicodeCharacter{2082}{\ensuremath{{}_2}}
\DeclareUnicodeCharacter{2083}{\ensuremath{{}_3}}
\DeclareUnicodeCharacter{2095}{\ensuremath{{}_h}}
\DeclareUnicodeCharacter{1D62}{\ensuremath{{}_i}}
\DeclareUnicodeCharacter{2C7C}{\ensuremath{{}_j}}
\DeclareUnicodeCharacter{2096}{\ensuremath{{}_k}}
\DeclareUnicodeCharacter{2097}{\ensuremath{{}_l}}
\DeclareUnicodeCharacter{2098}{\ensuremath{{}_m}}
\DeclareUnicodeCharacter{2099}{\ensuremath{{}_n}}
\DeclareUnicodeCharacter{2209}{\ensuremath{\notin}}
\DeclareUnicodeCharacter{27E6}{\ensuremath{\llbracket}}
\DeclareUnicodeCharacter{27E7}{\ensuremath{\rrbracket}}
\DeclareUnicodeCharacter{225F}{\ensuremath{\stackrel{?}{=}}}
\DeclareUnicodeCharacter{22A2}{\ensuremath{\vdash}}
\DeclareUnicodeCharacter{2218}{\ensuremath{\circ}}
\DeclareUnicodeCharacter{2032}{'}
\DeclareUnicodeCharacter{2227}{\ensuremath{\land}}
\DeclareUnicodeCharacter{2228}{\ensuremath{\vee}}

\definecolor{airforceblue}{rgb}{0.36, 0.54, 0.66}
\definecolor{brickred}{rgb}{0.8, 0.25, 0.33}
\definecolor{ao}{rgb}{0.0, 0.0, 1.0}
\definecolor{cobalt}{rgb}{0.0, 0.28, 0.67}
\definecolor{darkergreen}{rgb}{0,0.7,0.3}
\definecolor{magenta}{rgb}{1.0,0.0,1.0}

\FXRegisterAuthor{aa}{aaa}{AA}%
\FXRegisterAuthor{ms}{ams}{MS}%

\makeatletter
\providecommand*{\Dashv}{%
  \mathrel{%
    \mathpalette\@Dashv\vDash
  }%
}
\newcommand*{\@Dashv}[2]{%
  \reflectbox{$\m@th#1#2$}%
}
\makeatother
\makeatletter %
\def\arcr{\@arraycr}
\makeatother

\newcommand\ie{\emph{i.e.}, }
\newcommand\eg{\emph{e.g.}, }

\newcommand{\iLean}[1]{\mintinline{lean4}{#1}}

\definecolor{stringcolor}{HTML}{BA2121}

\makeatletter
\def\@parfont{\bfseries\itshape}
\makeatother

\newcommand{\subfigurespace}{0.2cm}

\newcommand{\osirismicro}[1]{\mathit{micro}~#1}
\newcommand{\HITree}{HITree\xspace}
\newcommand{\HITrees}{HITrees\xspace}

\newcommand{\ExampleLang}{ExampleLang\xspace}
\newcommand{\SimpLang}{$\lambda_{\mathit{par}, {\mathit{callcc}}}$\xspace}
\newcommand*{\valset}{\textsf{Val}}
\newcommand*{\exprset}{\textsf{Exp}}
\newcommand\langplus{\mathbin{\widehat{+}}}
\newcommand\langeq{\mathbin{\widehat{=}}}
\newcommand*{\alloc}{\texttt{ref}}
\newcommand*{\parcomp}[2]{#1 \mathop{||} #2}
\newcommand*{\callcc}[1]{\texttt{call/cc}(#1)}
\newcommand*{\throw}[1]{\texttt{throw}(#1)}
\newcommand*{\asrt}[1]{\texttt{assert}(#1)}
\newcommand*{\faa}[1]{\texttt{faa}(#1)}
\newcommand{\grayout}[1]{{\color{black!50} #1}}

\begin{document}

\title{HITrees: Higher-Order Interaction Trees}

\author{Amir Mohammad Fadaei Ayyam}
\orcid{0009-0005-5038-1619}
\affiliation{%
  \institution{Institute of Science and Technology (ISTA)}
  \city{Klosterneuburg}
  \country{Austria}
}
\affiliation{%
  \institution{Sharif University of Technology}
  \department{Computer Engineering Department}
  \city{Tehran}
  \country{Iran}
}
\email{am.fadaei@ce.sharif.edu}          %

\author[M.\ Sammler]{Michael Sammler}
\orcid{0000-0003-4591-743X}
\affiliation{%
  \institution{Institute of Science and Technology (ISTA)}
  \city{Klosterneuburg}
  \country{Austria}
}
\email{michael.sammler@ista.ac.at}

\begin{abstract}
Recent years have witnessed the rise of compositional semantics as a foundation for formal verification of complex systems. In particular, interaction trees have emerged as a popular denotational semantics.
Interaction trees achieve compositionality by providing a reusable library of effects.
However, their notion of effects does not support higher-order effects, \ie effects that take or return monadic computations.
Such effects are essential to model complex semantic features like parallel composition and call/cc.

We introduce \emph{Higher-Order Interaction Trees (HITrees)}, the first variant of interaction trees to support higher-order effects in a non-guarded type theory. HITrees accomplish this through two key techniques: first, by designing the notion of effects such that the fixpoints of effects with higher-order input can be expressed as inductive types inside the type theory; and
second, using defunctionalization to encode higher-order outputs into a first-order representation.
We implement HITrees in the Lean proof assistant, accompanied by a comprehensive library of effects including concurrency, recursion, and call/cc.
Furthermore, we provide two interpretations of HITrees, as state transition systems and as monadic programs.
To demonstrate the expressiveness of HITrees, we apply them to define the semantics of a language with parallel composition and call/cc.
 \end{abstract}



\maketitle

\section{Introduction}
\label{sec:intro}

Formal verification using machine-checked proofs has seen important advances in recent years, leading researchers to tackle more and more complex languages and systems~\cite{CertiKOS, seL4, SeKVM-SP, VellVM2}.
At the core of these verification projects are the formal semantics of the system that should be verified.
As these systems become more complex, their semantics also increases in complexity.
This has led to a rise of research on how to formally embed complex semantics into theorem provers~\cite{InteractionTrees, FreeSpec, GITrees, GITreesContext, Omnisemantics}.

A common method of embedding semantics into a theorem prover is to use \emph{operational semantics} (either big-step or small-step).
Operational semantics are very well understood (though there are still new, easier to use variants being discovered~\cite{Omnisemantics}) and widely used by verification projects~\cite{CompCert, IrisGroundUp}.
However, operational semantics have one important limitation: They can be challenging to compose modularly.
Operational semantics are usually defined as a monolithic inductive definition giving the transition system for the language in question, which can make it difficult to reuse components and constructs between different languages.

This lack of compositionality has led to a rise of research into more compositional methods of defining semantics via \emph{denotational semantics}.
A particularly active area of research on denotational semantics is based on variants of the free monad~\cite{DatatypesALaCarte}.
While free monads have a long tradition as denotational semantics~\cite{FreerMonad, McBrideTuringComplete, FreeSpec},
they started receiving significantly more attention for defining the semantics of programming languages with the arrival of \emph{interaction trees} (ITrees)~\cite{InteractionTrees}.
ITrees arose from the desire to encode the complex semantics of LLVM in a compositional way as part of the VellVM project~\cite{VellVM2}.
Since then, they have been picked up by a wide range of verification projects~\cite{CCR, DimSum, VellVM2, ProgramLogicsALaCarte, ITreeWeakMem}

The key insight of a denotational semantics based on ITrees (or a free monad in general) is
to build a compositional semantics around a set of \emph{uninterpreted effects}.
Such effects can include state, concurrency, non-determinism, and many more.
Using ITrees, one can build a complex denotational semantics using many different effects by composing reusable building blocks for individual effects.
This compositionality significantly simplifies the definition of complex semantics~\cite{VellVM2, ProgramLogicsALaCarte}.

While the notion of effects supported by ITrees and most variants of the free monad is quite general, there is one important class of effects that is not supported: \emph{Higher-order effects} that take a monadic computation as an input or return a monadic computation.
Such higher-order effects are useful to model complex higher-order languages.
As a concrete example, consider the recent formalization of OCaml by~\citet{Osiris}. Its monad $\osirismicro{A}$ exposes three different higher-order effects:%
\footnote{This presentation is simplified. For the full version, see \citet{Osiris}.}
\begin{align*}
  \mathit{par}~&(m_1 : \osirismicro{A})~(m_2 : \osirismicro{B}) : \osirismicro{(A \times B)} \\
  \mathit{try}~&(m : \osirismicro{A}) : \osirismicro{(A + \mathit{Exn})} \\
  \mathit{handle}~&(m : \osirismicro{A})~(m' : \mathit{Eff} \rightarrow \osirismicro{A}) : \osirismicro{A}
\end{align*}
The effect $\mathit{par}~m_1~m_2$ denotes the parallel composition of the monadic computations $m_1$ and $m_2$.
The effect $\mathit{try}~m$ catches exceptions $\mathit{Exn}$ raised by $m$.
Finally, the effect $\mathit{handle}~m~m'$ uses $m'$ to interpret the algebraic effect $\mathit{Eff}$ during the execution of $m$.
While this example shows the usefulness of higher-order effects, it has one important caveat: The $\osirismicro{A}$ monad is defined in a monolithic way---hard-coding the specific effects necessary for modeling OCaml.
While this works for \citet{Osiris}, this solution is not very satisfying.
By hard-coding the effects, one loses many of the benefits of the free monad---in particular, the ability to have a composable library of effects that can be reused across different languages.

This leads us to the core question of this paper:
\emph{How can one define a compositional, free-monad-based denotational semantics that supports higher-order effects?}

Answering this question is surprisingly challenging: None of the popular variants of the free monad, including ITrees~\cite{InteractionTrees} and FreeSpec~\cite{FreeSpec} support higher-order effects.
To our knowledge, the only variant of free monads supporting higher-order effects generically in a theorem prover is the GITree library by \citet{GITrees}.
GITrees support arbitrary higher-order effects with computations both as inputs and outputs of effects.
\citet{GITrees} show how GITrees can model complex effects like a higher-order store containing GITrees, or control effects like call/cc~\cite{GITreesContext}.
However, to achieve this expressiveness, GITrees have to pay a steep price:
Instead of defining the free monad as a normal (co-)inductive type in the ambient type theory of the proof assistant, GITrees are defined as the solution of a complex recursive domain equation inside a guarded type theory.
This means that working with GITrees requires working inside this guarded type theory, not the ambient type theory of the proof assistant.
In particular, one cannot use datatypes like pairs from the ambient type theory but has to define them in the guarded type theory (\eg via the church encoding) and then reason about them in the guarded type theory using step-indexing.

In this paper, we propose \emph{higher-order interaction trees (\HITrees)}, \emph{the first version of interaction trees supporting higher-order effects in a non-guarded type theory}.
In particular, we define \HITrees as an inductive%
\footnote{\autoref{sec:related-work} discusses why \HITrees are inductive instead of co-inductive.}
data type inside the Lean proof assistant and show that \HITrees support a large variety of higher-order effects.

To support higher-order effects, \HITrees leverage two techniques, one for handling higher-order inputs,%
\footnote{Technically, strictly positive occurrences of \HITrees in effects.}
one for higher-order outputs.%
\footnote{Technically, for negative occurrences of \HITrees in effects.}
Let us give a glimpse of the high-level ideas of these techniques here. (A detailed description can be found in \autoref{sec:semantics-example}.)

To support higher-order inputs, \HITrees adapt the notion of effects to avoid any dependency between the input and output types of the effect.
This enables computing the fixpoint of effects with higher-order inputs as an inductive type inside the ambient type theory of the proof assistant.
While the restriction that output types cannot depend on inputs may appear limiting, we demonstrate in \autoref{sec:semantics-example} how dependent effects can be recovered for \HITrees.

To support higher-order outputs, we use a variant of \emph{defunctionalization}~\cite{Defunctionalization}. Concretely, we observe that when an effect returns a \HITree, it represents some abstract computation whose purpose is to be run at some point.
This means that we can transform an effect with higher-order outputs to return an abstract identifier instead of a \HITree, together with exposing a separate effect to run the computation associated with the identifier.

Together, these techniques enable \HITrees to support a large variety of higher-order effects including the $\mathit{par}$ effect from above, a $\mathit{call/cc}$ effect that allows capturing the current continuation, and a fixpoint combinator $\mathit{rec}$.

We implement \HITrees in the Lean proof assistant~\cite{Lean4}.
Our implementation comes with a rich library of effects including state, non-determinism, concurrency, recursion, and call/cc.
Thanks to Lean's rich meta-programming features, our implementation of \HITrees provides custom commands for declaring new effects and taking the fixpoint of a higher-order effect.
To reason about and execute \HITrees, we develop two interpretations: One as state transition systems (inspired by \citet{ProgramLogicsALaCarte}) and one as an executable function into a suitably chosen monad.
Additionally, we show how to embed a language with parallel composition and call/cc into \HITrees.
Overall, we make the following contributions:
\paragraph{Contributions}
\begin{itemize}
\item \HITrees: a version of interaction trees supporting higher-order effects (\autoref{sec:gtree})
\item An implementation of \HITrees and their effects in Lean including meta-theory and commands for defining new effects and taking the fixpoint of effects
\item Two interpretations of \HITrees (\autoref{sec:interpretations}), one as state transition systems and one as executable monads.
\item A library of effects including state, non-determinism, concurrency, recursion, and call/cc. (\autoref{sec:effects})
\item A formalization of a language with concurrency and call/cc as a case study of \HITrees (\autoref{sec:case-study})
\end{itemize}
\autoref{sec:keyideas} provides an introduction to \HITrees,
and \autoref{sec:related-work} compares to related work.
Our Lean development is provided at \url{https://git.ista.ac.at/plv/hitrees}.
\section{\HITrees by Example}
\label{sec:keyideas}

\begin{figure}
  \centering
      \begin{align*}
        v \in \valset &\Coloneqq z \mid \ell \mid (v_1, v_2) \mid \Lam x. e \quad (z \in \mathbb Z, \ell \in \mathbb N)
        \\
        e \in \exprset &\Coloneqq v \mid x \mid e_1 \langplus e_2 \mid e.1 \mid e.2 \mid  \deref{e} \mid e_1 \gets e_2 \mid \\
        & \alloc(e) \mid \parcomp{e_1}{e_2} \mid e_1(e_2)
    \end{align*}
  \caption{\ExampleLang}
  \label{fig:examplelang}
\end{figure}

We introduce \HITrees via \ExampleLang (\autoref{fig:examplelang}), an untyped lambda calculus with integers $z$, pairs, a heap with locations $\ell$, and concurrency via parallel composition.

\subsection{Semantics of \ExampleLang}
\label{sec:semantics-example}

This section constructs a denotational semantics $\Sem{e}$ for \ExampleLang using \HITrees by gradually introducing the different effects of \ExampleLang.

\begin{figure}
  \centering
\begin{minted}{lean4}
structure Effect where
  I : Type
  O : Type

inductive HITree (E : Effect) (ρ : Type) where
  | pure (r : ρ)
  | impure (i : E.I) (k : E.O → HITree E ρ)
  | unreachable
\end{minted}

  \caption{(Preliminary) definition of effects and \HITrees}
  \label{fig:key:gtree}
\end{figure}

\paragraph{\#1 Failure: Introducing \HITrees}
We start with the semantics of addition $\Sem{e_1 \langplus e_2}$.
For this, we introduce our first effect: \emph{failure} (also called getting stuck, crashing, or undefined behavior).
Addition is only defined on integers, but not on lambda terms.
Thus, the semantics of addition first checks that the summands are integers and fail otherwise.
This \iLean{fail} operation is provided by the failure effect \iLean{FailE}, which we define using the \iLean{effect} command provided by our library:
\begin{minted}{lean4}
effect FailE where
  | fail : Unit → Empty
\end{minted}
This command declares the effect \iLean{FailE} with a single operation  \iLean{fail}.
Concretely, effects are a pair of an input type \iLean{I} and an output type \iLean{O}, represented by the \iLean{Effect} type in \autoref{fig:key:gtree}.%
\footnote{We omit universe parameters to not clutter the presentation.}
The \iLean{effect} command creates an \iLean{Effect} by generating an input type and an output type, each with one constructor per operation.
Concretely, using Lean's meta-programming, the \iLean{effect} command from above generates the following types:
\begin{minted}{lean4}
-- automatically generated by effect FailE
inductive FailE.I where | fail
inductive FailE.O where | fail (_ : Empty)
def FailE : Effect := ⟨FailE.I, FailE.O⟩
\end{minted}
\HITrees, shown in \autoref{fig:key:gtree},%
\footnote{This figure omits an intermediate definition, see \autoref{sec:gtree} for details.}
are a variant of interaction trees over \iLean{Effect}.
\HITrees have three constructors:%
\footnote{The names ``pure'' and ``impure'' are inspired by \citet{FreerMonad}.}
\iLean{pure r} to denote a pure computation resulting in a value \iLean{r} of type \iLean{ρ};
\iLean{impure i k} for an (impure) invocation of the effect with input \iLean{i} and continuation \iLean{k} that takes the result (of output type \iLean{E.O}) as the argument; and
\iLean{unreachable} as a denotation for impossible branches that cannot be ruled out statically.

The \iLean{effect FailE} command automatically lifts the \iLean{fail} operation to a function on \iLean{HITree} by generating a \iLean{fail} function:%
\footnote{This definition is simplified. We will come back to it later.}
\begin{minted}{lean4}
-- automatically generated by effect FailE
def fail : HITree FailE ρ :=
  .impure FailE.I.fail (λ x => x.1.elim)
\end{minted}

Now, finally everything is in place to define the denotation $\Sem{e_1 \langplus e_2}$.
First, we define the effect \iLean{ExE} of \ExampleLang.
For now, it contains only \iLean{FailE}, but we will extend it later.
\begin{minted}{lean4}
def ExE := FailE
\end{minted}
Then, we define a function \iLean{intOrFail} to checks if a value is an integer and fail otherwise.
\begin{minted}{lean4}
def Val.intOrFail : Val → HITree ExE Int :=
  | .int z => return z
  | _ => fail
\end{minted}
Then, we can define $\Sem{e_1 \langplus e_2}$ (in Lean: \iLean{(.plus e₁ e₂).denote}):
\begin{minted}{lean4}
def Exp.denote : Exp → HITree ExE Val := -- e.denote is $\Sem{e}$
  | .plus e₁ e₂ => do -- .plus e₁ e₂ is $e_1 \langplus e_2$
    let v₁ ← e₁.denote; let v₂ ← e₂.denote
    return .int ((← v₁.intOrFail) + (← v₂.intOrFail)) ...
\end{minted}
\HITrees form a monad (see \autoref{sec:definition-of-gtree}) and thus we can use Lean's \iLean{do} notation~\cite{Lean4Do} for monads.
This notation elaborates into a sequence of \HITree binds where each left arrow \iLean{←} corresponds to one bind.

\paragraph{\#2 State: Unreachable}
Let us now continue with the denotations of the next expressions of \ExampleLang: loading from the heap ($\deref{e}$) and storing to the heap ($e_1 \gets e_2$).
To model the heap, we introduce the \iLean{StateE} effect.
\begin{minted}{lean4}
effect StateE (α : Type) where
  | get : Unit → α | set : α → Unit
\end{minted}
This effect is parametrized by the type $\alpha$ of the state. It provides two operations, \iLean{get} and \iLean{set}, for retrieving resp. updating the state.
As before, the \iLean{effect} command automatically generates the input and output types from this description:
\begin{minted}{lean4}
-- automatically generated by effect StateE
inductive StateE.I (α : Type) where
  | get         | set (_ : α)
inductive StateE.O (α : Type) where
  | get (_ : α) | set
def StateE (α : Type) : Effect := ⟨StateE.I α, StateE.O α⟩
\end{minted}
However, when we try to use this effect with \iLean{.impure}, we run into an issue. Intuitively, an input \iLean{StateE.I.get} should always result in an output \iLean{StateE.O.get}, but nothing in the type system enforces this.
ITrees address this issue by using a more dependent notion of effects where each input comes with the type of its output. However, this dependency prevents higher-order effects (due to a negative occurrence of the effect type, as we will see later in this section).
To enable higher-order effects, \HITrees avoid a dependency of the output type on the input. Instead, we implement this dependency using a dynamic check.
Concretely, the \iLean{effect} command generates the following function for \iLean{get}.%
\footnote{The version here is preliminary, it will be slightly adapted later.}
\begin{minted}{lean4}
-- automatically generated by effect StateE
def get {α : Type} : HITree (StateE α) α :=
  .impure (StateE.I.get) λ
    | StateE.O.get o => .pure o
    | _ => .unreachable
\end{minted}
This function uses a dynamic check to inspect the output of the \iLean{get} operation. If it is \iLean{StateE.O.get o}, \iLean{get} returns \iLean{o}. Otherwise, it discards the execution using \iLean{.unreachable}.
Thanks to this automatically generated dynamic check, the user can use \iLean{get} with expected type \iLean{HITree (StateE α) α}.

\begin{figure}
  \centering
\begin{minted}{lean4}
def SumE (E₁ : Effect) (E₂ : Effect) : Effect
  := ⟨E₁.I ⊕ E₂.I, E₁.O ⊕ E₂.O⟩
infixr:50 " ⊕ₑ " => SumE

class Subeffect (E₁ : Effect) (E₂ : Effect) where
  mapI : E₁.I → E₂.I
  mapO? : E₂.O → Option E₁.O
infix:20 " -< " => Subeffect

instance {E} : E -< E where
  mapI i := i
  mapO? o := some o

instance {E₁ E₂ E'} [sub : E₁ -< E₂] : E₁ -< (E₂ ⊕ₑ E') where
  mapI i := .inl (sub.mapI i)
  mapO? := λ
  | .inl o => sub.mapO? o
  | .inr _ => none

instance {E₁ E₂ E'} [sub : E₁ -< E₂] : E₁ -< (E' ⊕ₑ E₂) where
  ... -- analogous to the previous instance

def trigger {E₁ E₂} [sub : E₁ -< E₂] (i : E₁.I) : HITree E₂ E₁.O
  := .impure (sub.mapI i) λ x => match sub.mapO? x with
       | some x => .pure x
       | none => .unreachable
\end{minted}

  \caption{Effect composition}
  \label{fig:key:comp}
\end{figure}

\paragraph{Effect composition}
So far, we have only seen \HITrees using single effects, but our denotation $\Sem{e}$ requires both the \iLean{FailE} and the \iLean{StateE} effect.
To combine effects, \HITrees provide the infrastructure shown in \autoref{fig:key:comp}.
The sum \iLean{E₁ ⊕ₑ E₂} of \iLean{E₁} and \iLean{E₂} is defined as the sum of the input type and of the output type.
With this, we can define \iLean{ExE} as the sum of \iLean{FailE} and \iLean{StateE} (using a treemap with extensional equality to denote the heap as a map from location to values).
\begin{minted}{lean4}
def ExE := FailE ⊕ₑ StateE (ExtTreeMap Loc Val)
\end{minted}
When trying to use \iLean{get} from above with \iLean{ExE}, we run into an issue: \iLean{get} is specific to the effect \iLean{StateE α} and thus does not work for \iLean{ExE}.
To address this issue, we introduce the notion of \emph{subeffects}~\cite{FreerMonad, InteractionTrees}, captured by the \iLean{Subeffect} typeclass in \autoref{fig:key:comp}.
\iLean{E₁} is a subeffect of \iLean{E₂} (written \iLean{E₁ -< E₂}) if there is a map \iLean{mapI} from the inputs of \iLean{E₁} to the inputs of \iLean{E₂} and a map \iLean{mapO?} from the outputs of \iLean{E₂} to the outputs of \iLean{E₁}.
Since \iLean{E₂} represents a ``bigger'' effect than \iLean{E₁}, \iLean{mapO?} can fail (by returning \iLean{none}) for outputs of \iLean{E₂} that do not correspond to outputs of \iLean{E₁}.
\iLean{E₁ -< E₂} is reflexive and (subeffects of) \iLean{E₁} and \iLean{E₂} are subeffects of \iLean{E₁ ⊕ₑ E₂}:

The \iLean{trigger} function (\autoref{fig:key:comp}) invokes a subeffect \iLean{E₁} in a larger effect \iLean{E₂}.
\iLean{trigger} uses \iLean{mapI} to inject \iLean{i} into \iLean{E₂} and \iLean{mapO?} to project out the result.
The \iLean{none} case of \iLean{mapO?} is \iLean{unreachable} since the input of the \iLean{.impure} was an operation of \iLean{E₁}.

Using \iLean{trigger}, we can define \iLean{get} in a way that automatically works for arbitrary supereffects.
\begin{minted}{lean4}
-- automatically generated by effect StateE
def get {α : Type} {E} [StateE α -< E] : HITree E α :=
  trigger (StateE.I.get).bind λ
    | StateE.O.get o => .pure o
    | _ => .unreachable
\end{minted}

This allows us to define $\Sem{\deref{e}}$ ($\Sem{e_1 \gets e_2}$ is analogous):
\begin{minted}{lean4}
def Exp.denote : Exp → HITree ExE Val :=
  | .deref e => do -- .deref e is $\deref{e}$
    let v ← e.denote;
    let h ← get; let l ← v₁.locOrFail;
    let some r := h[l]? | fail
    return r ...
\end{minted}

\paragraph{\#3 Demonic choice: Dependent effects}
Let us now turn to the last remaining heap operation: allocation $\alloc(e_1)$. Allocation is usually modeled using \emph{demonic non-determinism}, \ie the allocator picks an arbitrary location that is not allocated yet.
To represent demonic non-determinism, we introduce a new effect \iLean{DemonicE}. How should we define \iLean{DemonicE}?
\citet{ProgramLogicsALaCarte} define demonic choice by taking a type $\alpha$ as an input and returning an element of type $\alpha$:
\begin{minted}{lean4}
effect DemonicE where -- try 1
  | choose : (α : Type) → α
\end{minted}
However, this definition is rejected since our \iLean{Effect} type does not support the output type to depend on the input.%
\footnote{An additional problem is that this \iLean{DemonicE} must live in a higher universe than $\alpha$. Combining this \iLean{DemonicE} with effects in lower universes requires manual universe lifting since Lean lacks universe cumulativity.}
We can mitigate this problem by making $\alpha$ a parameter of \iLean{DemonicE}:
\begin{minted}{lean4}
effect DemonicE (α : Type) where -- try 2
  | choose : Unit → α
\end{minted}
This definition works, but it is quite restrictive as one needs to fix the types of demonic choice statically. This is an issue for $\alloc(e)$: It should pick an arbitrary location \emph{that is not allocated on the current heap}. Since the current heap is not available statically, we cannot refer to it in $\alpha$.

Thus, we arrive at our final definition of \iLean{DemonicE}:
\begin{minted}{lean4}
effect DemonicE (α : Type) where -- final
  | choose : (p : α → Prop) → [DecidablePred p] → α
\end{minted}
The idea is to fix $\alpha$ statically, but allow dynamically picking a decidable predicate $p$ that restricts the possible values of the demonic choice.
This allows us to define a \iLean{choose} function that guarantees that the resulting value satisfies $p$.
\begin{minted}{lean4}
def choose {α : Type} (p : α → Prop) [DecidablePred p]
  {E} [DemonicE α -< E] : HITree E (x : α // p x) :=
  trigger (Demonic.I.choose p).bind λ o =>
    if h : p o.1 then return ⟨o.1, h⟩ else .unreachable
\end{minted}
With this, we can define $\Sem{\alloc(e)}$ as follows:
\begin{minted}{lean4}
def Exp.denote : Exp → HITree ExE Val :=
  | .ref e => do
    let v ← e.denote; let h ← get;
    let l ← choose (λ l => l ∉ h); set (h.insert l.1 v)
    return r ...
\end{minted}

This example of \iLean{DemonicE} shows how \HITrees deal with dependencies between inputs and outputs of effects. The strict separation of input and output in \iLean{Effect} rules out direct dependencies of the output type on the input. This prevents some effects like try 1 of \iLean{DemonicE} that are possible in more dependent encodings like interaction trees.
However, one can still have an operation where the type of the output depends on the input, as long as one can provide a non-dependent output type ($\alpha$ in \iLean{DemonicE}) that can be decidably cast into the dependent output type (\iLean{x : α // p x} in \iLean{DemonicE}).

\paragraph{\#4 Parallel: Higher-order effects}
While the previous section showed the cost of using a non-dependent notion of effects, we will now discuss the payoff: higher-order effects.
For this, consider parallel composition $\parcomp{e_1}{e_2}$.
To model this construct, we introduce a new effect \iLean{ConcE} for concurrency.
\begin{minted}{lean4}
effect ConcE (α : Type) (E : Effect) where
  | par : (t₁ t₂ : HITree E Empty) → (α × α)
  | kill : α → Empty
  | yield : Unit → Unit
\end{minted}
The main operation of this effect is the parallel composition \iLean{par} that takes two \HITrees, runs them in parallel, and returns a pair of the results.
The result of a thread is given to the \iLean{kill} operation that terminates the current thread.
Additionally, the effect exposes a \iLean{yield} operation for switching to another thread. We adapt $\Sem{e}$ such that it inserts \iLean{yield} at the appropriate points, following \citet{ProgramLogicsALaCarte}.

The challenge of \iLean{ConcE} is that \iLean{par} takes \HITrees as arguments and these \HITrees should be able to use parallel composition themselves. In particular, the effect parameter \iLean{E} of \iLean{ConcE} should contain \iLean{ConcE}, creating a recursive loop.
Concretely, we would like to define \iLean{ExE} as follows:
\begin{minted}{lean4}
-- error: fail to show termination
def ExE := ConcE Val ExE ⊕ₑ FailE ⊕ₑ ...
\end{minted}
However, Lean rejects this recursive usage of \iLean{ExE} since it cannot show termination of this recursive definition.

To find the fixpoint of higher-order effects like \iLean{ExE}, \HITrees provide the \iLean{fix_effect} command.
\begin{minted}{lean4}
fix_effect ExE := ConcE Val ExE ⊕ₑ FailE ⊕ₑ ...
\end{minted}
This command defines \iLean{ExE} as a recursive effect that is isomorphic to the rejected version of \iLean{ExE} above.

Using this \iLean{ExE}, we can define $\Sem{\parcomp{e_1}{e_2}}$ as follows:
\begin{minted}{lean4}
def Exp.denote : Exp → HITree ExE Val :=
 | .par e₁ e₂ => do -- .par e₁ e₂ is $\parcomp{e_1}{e_2}$
  let (v₁, v₂) ← par (e₁.denote >>= kill) (e₂.denote >>= kill)
  return .pair v₁ v₂ ...
\end{minted}
Note that it is crucial that \iLean{par} takes \iLean{HITree ExE Empty} as the argument to support nested parallel composition $\Sem{\parcomp{\parcomp{e_1}{e_2}}{e_3}}$.

Let us now have a look under the hood of \iLean{fix_effect}.
Our key observation is that Lean (and CIC in general) provides a native way to take the fixpoint of a type: the \iLean{inductive} command.
Inductive types can contain recursive occurrences of themselves, as long as all occurrences are \emph{strictly positive}~\cite{InductivelyDefinedTypes}. This means the recursive occurrences cannot be on the left-hand side of an arrow and (in Lean) are only allowed to appear either directly or in arguments to other inductive types.%
\footnote{See \url{https://lean-lang.org/doc/reference/4.22.0/The-Type-System/Inductive-Types/\#--tech-term-Nested-inductive-types} for details.}

The following shows the code that \iLean{fix_effect ExE} generates to find the fixpoint of \iLean{ExE} using the \iLean{inductive} command.%
\footnote{The inductive type here is simplified (see \autoref{sec:fixpoints-of-effects} for details).}
\begin{minted}{lean4}
-- automatically generated from fix_effect ExE
def ExE.O := ConcE.O Val ⊕ FailE.O ⊕ ...
inductive ExE.I where
  | closed (_ : ConcE.I Val ⟨ExE.I, ExE.O⟩ ⊕ FailE.I ⊕ ...)
def ExE : Effect := ⟨ExE.I, ExE.O⟩
\end{minted}
It splits \iLean{ExE} into the output type \iLean{ExE.O} and input type \iLean{ExE.I}.
Only the input type occurs in a strictly positive position in the \iLean{HITree} type and thus the output type must not contain a recursive occurrence of \iLean{ExE}.
The input type \iLean{ExE.I} is defined as an inductive type that recursively appears as an argument to \iLean{ConcE.I}.
This works since the \iLean{Effect} type is designed such that the output type does not depend on the input---if the output type would depend on the input, the input type would need to appear as part of the output type in a not strictly-positive position and Lean would reject the inductive type.

\paragraph{\#5 Recursion: Negative occurrences via defunctionalization}
Let us now see how \HITrees can support higher-order operations where the recursive occurrence is in a \emph{negative} position.
For this, we consider the semantics of function application $\Sem{e_1(e_2)}$.
Here the challenge is that function calls can introduce non-termination and we need a way to represent this non-termination in our semantic domain of \HITrees.
One option is to use a co-inductive definition based on the Delay monad~\cite{DelayMonad} as used by ITrees.
However, since Lean does not have native support for co-inductive types, we cannot use this option.
Instead, we use a higher-order effect for recursion to represent non-termination.
Concretely, our goal is to have a recursion combinator \iLean{rec} on \HITrees:
\begin{minted}{lean4}
rec : ((α→HITree E β) → (α→HITree E β)) → α→HITree E β
\end{minted}
A naive idea would be to define \iLean{rec} directly as an effect:
\begin{minted}{lean4}
-- error: not strictly positive occurrence of E
effect RecE (α β : Type) (E : Effect) where
| fix : (_ : (α→HITree E β) → α→HITree E β) (_ : α) → β
\end{minted}
However, this approach does not work since the first occurrence of \iLean{E} is in a not in a strictly positive position.
To resolve this issue, we use \emph{defunctionalization}~\cite{Defunctionalization}.
Concretely, we introduce a set of abstract identifiers \iLean{FId} that represent fixpoints with an operation \iLean{call : FId→α→HITree E β} that invokes the fixpoint represented by the \iLean{FId}:
\begin{minted}{lean4}
effect RecE (α β : Type) (E : Effect) where
  | fix : (_ : FId → α → HITree E β) (_ : α) → β
  | call : (_ : FId) (_ : α) → β
\end{minted}
This definition avoids the not strictly positive occurrence of \iLean{E} by using \iLean{FId}.
We can obtain \iLean{rec} by combining \iLean{fix} and \iLean{call}.
\begin{minted}{lean4}
def rec [RecE α β E -< E]
   (f : (α → HITree E β) → α → HITree E β) :
 α → HITree E β := fix λ fid => f (call fid)
\end{minted}

The recursion combinator \iLean{rec} lets us define $\Sem{e_1(e_2)}$:
\begin{minted}{lean4}
def Exp.denoteRaw (denote : Exp → HITree ExE Val)
  : Exp → HITree ExE Val :=
  | .app e₁ e₂ => do -- .app e₁ e₂ is $e_1(e_2)$
    let v₁ ← e₁.denoteRaw; let v₂ ← e₂.denoteRaw;
    let ⟨x, e⟩ ← v₁.toRec! -- turn v₁ into λ x. e
    denote (e.subst x v₂) -- ⟦e[v₂/x]⟧
...
def Exp.denote : Exp → HITree ExE Val := rec Exp.denoteRaw
\end{minted}
This definition uses the fixpoint \iLean{denote} provided by \iLean{rec} to recursively call \iLean{denoteRaw} on $e[v_2/x]$. Directly calling \iLean{denoteRaw} would be rejected by Lean since the call is not structurally recursive.
Overall, this shows how defunctionalization can encode higher-order effects with not strictly positive occurrences.

\subsection{Interpreting \HITrees}
\label{sec:interpreting-hitrees}

We have seen how \HITrees can give a denotational semantics to \ExampleLang.
Let us now see what we can do with such a semantics.
We focus on two aspects: \emph{execution} and \emph{reasoning}.

\paragraph{Executing \HITrees}
To test a formal semantics, it is useful to be able to execute it.
\HITrees can be executed by interpreting them into a suitable Lean \iLean{Monad}, which can then be compiled and executed like normal Lean code.
Concretely, each effect can provide a \emph{monad handler} that describes how to interpret the effect into a monad.
For example, the monad handler for the \iLean{ConcE} effect interprets the effect into a state monad that keeps track of all running threads and implements a scheduler that picks the next thread on \iLean{yield}.
\autoref{sec:eval} explains this in more detail.

\paragraph{Reasoning about \HITrees}
To reason about \HITrees, we implement the technique of \citet{ProgramLogicsALaCarte} to turn interaction trees into state machines.
In a nutshell, this means that each effect defines a small-step semantics, which can then be lifted to a big-step semantics $t \Downarrow v$ on \HITrees.
This big-step semantics allows us to prove that certain (problematic) executions exist.
For example, we can show that a race condition can cause the following program to return 1:
\[
  \Sem{\Let x = \alloc(0) in (\parcomp{x \gets \deref{x} + 1}{x \gets \deref{x} + 1}); \deref{x}} \Downarrow 1
\]
\autoref{sec:exec} discusses the state machine semantics in more detail and \autoref{sec:case-study} shows more examples.
In the future, it would be interesting to implement more powerful reasoning mechanisms for \HITrees, like the program logic of  \citet{ProgramLogicsALaCarte}.

\section{\HITrees in Detail}
\label{sec:gtree}
This section presents the definition of \HITrees (\autoref{sec:definition-of-gtree}) and the \iLean{effect} (\autoref{sec:effect-command}) and \iLean{fix_effect} (\autoref{sec:fixpoints-of-effects}) commands in detail.

\subsection{Definition of \HITrees}
\label{sec:definition-of-gtree}

\begin{figure}
  \centering

\begin{minted}{lean4}
inductive HITree.Raw (I O ρ) where
  | pure (r : ρ)
  | impure (i : I) (k : O → HITree.Raw I O ρ)
  | unreachable

def HITree (E : Effect) := HITree.Raw E.I E.O
-- ... lift pure, impure, unreachable to HITree ...

def HITree.bind (t : HITree E ρ) (f : ρ → HITree E ρ')
  := match t with
    | .unreachable => unreachable
    | .pure r => f r
    | .impure i k => impure i (λ x => bind (k x) f)

instance : Monad (HITree E) where
  pure := HITree.pure
  bind := HITree.bind
instance : LawfulMonad (HITree E) where ...
\end{minted}
  \caption{Full definition of \HITrees}
  \label{fig:gtree-full}
\end{figure}

\autoref{fig:gtree-full} shows the full definition of \HITrees.
This definition differs from \autoref{fig:key:gtree} by introducing the intermediate type \iLean{HITree.Raw}.
This type is necessary due to a problem we glanced over in \autoref{sec:keyideas}: The definition of \iLean{HITree E ρ} from \autoref{fig:key:gtree} actually contains the type of inputs \iLean{E.I} in a negative position: \iLean{E.I} it is contained in \iLean{E}, which appears in \iLean{E.O} that occurs in a negative position in \iLean{impure}.
This negative occurrence prevents taking the fixpoint of higher-order effects.

To resolve this issue, we introduce \iLean{HITree.Raw}. It ensures that the input type occurs only in strictly positive positions by separating the input type \iLean{I} from the output type \iLean{O}.
We obtain \iLean{HITree} as a straightforward definition on top of \iLean{HITree.Raw} and we lift \iLean{pure}, \iLean{impure}, and \iLean{unreachable} to \iLean{HITree}.
Users only interact with \iLean{HITree}. The only place where \iLean{HITree.Raw} appears is in the definition of higher-order effects.
This is necessary since higher-order effects are part of the fixpoint and thus must only contain the recursive occurrence in strictly positive positions. The \iLean{effect} command automatically handles this by desugaring effects using \iLean{HITree} to use \iLean{HITree.Raw}.

\HITrees provide a \iLean{bind} function (\autoref{fig:gtree-full}) and inhabit the \iLean{Monad} and \iLean{LawfulMonad} interfaces. The first instance enables us to use Lean's \iLean{do} notation to define \HITrees
and the second instance automatically registers the monad laws with Lean's simplifier tactic \iLean{simp}.
We reuse the \iLean{trigger} function from \autoref{fig:key:comp} for \iLean{HITree} with the definition of \HITree in \autoref{fig:gtree-full}.

\subsection{Defining Effects}
\label{sec:effect-command}
This section discusses the \iLean{effect} command for defining new effects and how it works under the hood.

\begin{figure}
  \centering
\begin{subfigure}{\columnwidth}
\begin{minted}{lean4}
effect Eff p₁ ... pᵢ where
  | op₁ : (x₁ : ι^$_{1,1}$^) → ... → (x^$_{m_1}$^ : ι^$_{1,m_1}$^) → o₁
  ...
  | opₙ : (x₁ : ι^$_{n,1}$^) → ... → (x^$_{m_n}$^ : ι^$_{n,m_n}$^) → oₙ
\end{minted}
  \caption{Syntax of \iLean{effect}.}
  \label{fig:effect:syntax}
\end{subfigure}

\begin{subfigure}{\columnwidth}
\vspace{\subfigurespace}
\begin{minted}{lean4}
inductive Eff.I p₁ ... pₖ where
  | op₁ (x₁ : ι^$_{1,1}$^) ... (x^$_{m_1}$^ : ι^$_{1,m_1}$^)
  ...
  | opₙ (xₙ : ι^$_{n,1}$^) ... (x^$_{m_n}$^ : ι^$_{n,m_n}$^)
inductive Eff.O p₁ ... pₖ where
  | op₁ (_ : o₁) | ... | opₙ (_ : oₙ)
def Eff p₁ ... pₖ : Effect :=
  ⟨Eff.I p₁ ... pₖ, Eff.O p₁ ... pₖ⟩

unif_hint (E : Effect) p₁ ... pₖ
  where E ≟ Eff ⊢ E.I ≟ Eff.I p₁ ... pₖ
unif_hint (E : Effect) p₁ ... pₖ
  where E ≟ Eff ⊢ E.O ≟ Eff.O p₁ ... pₖ

-- per operation opᵢ
def Eff.opᵢ {E} p₁ ... pₖ [Eff p₁ ... pₖ -< E]
            (x₁ : ι^$_{i,1}$^) ... (x^$_{m_i}$^ : ι^$_{i,m_i}$^) : HITree E oᵢ :=
  (trigger (.I.opᵢ p₁ ... pₖ x₁ ... x^$_{m_i}$^)).bind λ o =>
    if let .O.opᵢ o' := o then .pure o' else .unreachable
\end{minted}
  \caption{Desugaring of \iLean{effect}.}
  \label{fig:effect:generated}
\end{subfigure}
  \caption{The \iLean{effect} command.}
  \label{fig:effect}
\end{figure}

The syntax for declaring new effects is shown in \autoref{fig:effect:syntax}.
The \iLean{effect} command first expects the name of the effect \iLean{Eff} and its $k$ parameters \iLean{p₁, ..., pₖ}.
These parameters follow the same syntax as regular binders in Lean, supporting explicit, implicit and typeclass parameters.
Then follow the $n$ operations \iLean{op₁, ..., opₙ}, where each operation \iLean{opᵢ} has $m_i$ inputs of types \iLean{ι^$_{i,1}$^, ..., ι^$_{i,m_i}$^} respectively, and an output of type \iLean{oᵢ}.
The syntax of these operations follows the syntax of constructors of an inductive type, except that the output type \iLean{o₁} cannot depend on any of the input binders \iLean{x^$_j$^}.

Let us now see how the \iLean{effect} command automatically generates the corresponding \iLean{Effect}. Concretely, the \iLean{effect} command desugars to the Lean code shown in \autoref{fig:effect:generated}.

First, the \iLean{effect} command defines the input and output types of the effect as the inductive types \iLean{Eff.I} and \iLean{Eff.O} and defines \iLean{Eff} by combining them.
The input and output types each contain one constructor per operation with the corresponding inputs resp. outputs as parameters.

Next, the \iLean{effect} command generates two unification hints via the \iLean{unif_hint} command.
These hints guide Lean's unification when solving constraints of the form \iLean{?E.I ≟ Eff.I} and \iLean{?E.O ≟ Eff.O} by telling it to instantiate the unification variable \iLean{?E} with \iLean{Eff}.
For example, these hints allow Lean to infer that \iLean{HITree.impure (Eff.I.opᵢ ...) λ _ => return 0} has type \iLean{HITree (Eff ...) Nat} without  specifying \iLean{(E:=Eff ...)}.

Finally, for each operation \iLean{opᵢ} the \iLean{effect} command generates a specialized trigger function \iLean{Eff.opᵢ}.
This function casts the output to the right type by checking if the output corresponds to the same operation as the input and otherwise discarding the execution as unreachable (see also \autoref{sec:semantics-example}).

\paragraph{Special cases}
There are three special cases:
First, if the type of a parameter is \iLean{Unit}, the parameter is omitted (see \iLean{get} in \autoref{sec:semantics-example}).
Second, if the output type \iLean{oᵢ} is \iLean{Empty}, the trigger function \iLean{Eff.opᵢ} will be automatically generalized from \iLean{HITree E Empty} to \iLean{HITree E ρ} (for arbitrary \iLean{ρ}) by eliminating the \iLean{Empty} (see \iLean{fail} in \autoref{sec:semantics-example}).
Third, if a type of an input parameter \iLean{ι^${_{i,j}}$^} contains \iLean{HITree pₗ _}, \iLean{HITree pₗ _} is converted to \iLean{HITree.Raw pₗI pₗO _} in \iLean{ι^${_{i,j}}$^} and the effect parameter \iLean{pₗ} is replaced with two parameters \iLean{pᵢI} and \iLean{pᵢO} (updating the other definitions as necessary).
This desugaring of \iLean{HITree} to \iLean{HITree.Raw} is necessary to avoid not strictly positive occurrences as discussed in \autoref{sec:definition-of-gtree}.

\subsection{Fixpoints of Effects}
\label{sec:fixpoints-of-effects}
\begin{figure}
  \centering
\begin{subfigure}{\columnwidth}
\begin{minted}{lean4}
fix_effect Eff := E₁ ⊕ₑ ... ⊕ₑ Eₙ
\end{minted}
  \caption{Syntax of \iLean{fix_effect}.}
  \label{fig:fix:syntax}
\end{subfigure}

\begin{subfigure}{\columnwidth}
\vspace{\subfigurespace}
\begin{minted}{lean4}
def Eff.Pre (Eff : Effect) := E₁ ⊕ₑ ... ⊕ₑ Eₙ

-- EᵢO is the reduced form of Eᵢ.O
def Eff.O := E₁O ⊕ ... ⊕ EₙO

-- E$'_i$ is Eᵢ[⟨Eff.I, Eff.O⟩/Eff] and E$'_i$I
-- is the reduced form of E$'_i$.I
inductive Eff.I where
  | closed (_ : E^$'_1$^I ⊕ ... ⊕ E^$'_n$^I)

def Eff : Effect := ⟨Eff.I, Eff.O⟩

instance : UnfoldEffect Eff (Eff.Pre Eff) where
  -- body omitted
\end{minted}
  \caption{Desugaring of \iLean{fix_effect}.}
  \label{fig:fix:generated}
\end{subfigure}

\begin{subfigure}{\columnwidth}
\vspace{\subfigurespace}
\begin{minted}{lean4}
class UnfoldEffect (E : Effect) (EPre : outParam Effect) where
  injI : E.I → EPre.I
  invI : EPre.I → E.I
  injI_invI : ∀ i, injI (invI i) = i
  invI_injI : ∀ i, invI (injI i) = i
  out_eq : EPre.O = E.O

instance {E EPre E₁} [UnfoldEffect E EPre] [sub : E₁ -< EPre]
    : E₁ -< E where
  mapI i := UnfoldEffect.injI (sub.mapI i)
  mapO? o := sub.mapO? (Eq.mpr UnfoldEffect.out_eq o)
\end{minted}
  \caption{Unfolding effects.}
  \label{fig:fix:unfold}
\end{subfigure}

\caption{The \iLean{fix_effect} command.}
  \label{fig:fix}
\end{figure}
\autoref{fig:fix:syntax} shows the syntax of the \iLean{fix_effect} command along with the definitions it generates.
\iLean{fix_effect} takes the name \iLean{Eff} of the recursively defined effect and the body of the fixpoint. The body is a list of effects \iLean{E₁ ... Eₙ} combined using \iLean{⊕ₑ}. These effects can mention \iLean{Eff} as the recursive occurrence.

The \iLean{fix_effect} command generates several definitions shown in \autoref{fig:fix:generated}:
First, it defines \iLean{Eff.Pre} as the version of \iLean{Eff} that takes the recursive occurrence as an argument.

Second, the output type \iLean{Eff.O} is defined as the output type of the effect \iLean{E₁ ⊕ₑ ... ⊕ₑ Eₙ}.
To avoid a spurious dependency on the recursive occurrence \iLean{Eff}, we first reduce \iLean{(E₁ ⊕ₑ ... ⊕ₑ Eₙ).O} to \iLean{E₁O ⊕ ... ⊕ EₙO} where \iLean{EᵢO} is the output type of \iLean{Eᵢ}. (Recall that the output type cannot contain a recursive occurrence since it appears in a negative position in \iLean{HITree.Raw}.)
In case the \iLean{fix_effect} command detects a recursive occurrence in \iLean{Eff.O} or if it cannot perform the reduction, it emits an error.

Finally, we reach the heart of the \iLean{fix_effect} command: The definition of \iLean{Eff.I} as an inductive type.
Intuitively, this inductive type should contain a single field \iLean{(E₁ ⊕ₑ ... ⊕ₑ Eₙ).I} with the recursive occurrence replaced by \iLean{⟨Eff.I, Eff.O⟩}.
However, we need to be careful to satisfy all the conditions that Lean places on inductive types.
First, we must ensure that all occurrences of \iLean{Eff.I} are strictly positive.
As described in \autoref{sec:definition-of-gtree}, the \iLean{HITree.Raw} type is carefully set up to avoid not strictly positive occurrences.
Second, we need to ensure that \iLean{Eff.I} only appears as arguments to type constructors of inductive types.
In particular, we cannot use \iLean{(E₁ ⊕ₑ ... ⊕ₑ Eₙ).I} directly (since the projection \iLean{.I} is not a type constructor of an inductive type) but need to reduce it to expose input types of the effects \iLean{E₁I ⊕ ... ⊕ EₙI}.
Since the \iLean{effect} command generates the input types as inductive types, we know that \iLean{EᵢI} will be an inductive type.
(The reduction is also important to avoid spurious negative occurrences.)
Additionally, recall that the \iLean{effect} command (\autoref{sec:effect-command}) expanded \iLean{HITree E ρ} to \iLean{HITree.Raw EI EO ρ}, avoiding the use of the projection \iLean{.I}.
All of this together ensures that the generated inductive type is accepted by Lean.

As a concrete example, the input of the \iLean{ExE} effect from \autoref{sec:semantics-example} is defined as follows:
\begin{minted}{lean4}
fix_effect ExE := ConcE Val ExE ⊕ₑ Fail
-- generates the following:
inductive ExE.I where
  | closed (_ : ConcE.I Val ExE.I ExE.O ⊕ FailE.I)
\end{minted}
Note \iLean{ConcE} takes \iLean{ExE.I} and \iLean{ExE.O} as separate arguments (as described in \autoref{sec:effect-command}). This makes Lean accept this definition---in contrast to the simplified version shown in \autoref{sec:semantics-example} which would be rejected since the recursive occurrence of \iLean{ExE.I} appears under the constructor \iLean{⟨Eff.I, Eff.O⟩}, which is not a type constructor of an inductive type.

\iLean{Eff} is defined by combining \iLean{Eff.I} and \iLean{Eff.O}.

\paragraph{Unfolding recursive effects}
How can we work with the recursive effects defined via \iLean{fix_effect}?
This is where the \iLean{UnfoldEffect} typeclass comes into play, shown in \autoref{fig:fix:unfold}.
Intuitively, \iLean{UnfoldEffect E EPre} states that the recursive effect \iLean{E} is isomorphic to its unfolding \iLean{EPre}.
The \iLean{fix_effect} command generates an instance \iLean{UnfoldEffect Eff (Eff.Pre Eff)}.

\iLean{UnfoldEffect} is used to automatically lift properties of \iLean{EPre} to \iLean{E}. For example, \autoref{fig:fix:unfold} shows that if \iLean{E₁} is a subeffect of \iLean{EPre}, it is also a subeffect of \iLean{E}.
We already used this implicitly in \autoref{sec:semantics-example} when using operations of \iLean{ConcE} in \iLean{HITree ExE Val}.
Note that \iLean{EPre} is an \iLean{outParam} of \iLean{UnfoldEffect}.
This allows typeclass synthesis to determine \iLean{EPre} from \iLean{E}.

\section{Interpretations}
\label{sec:interpretations}
This section presents two interpretations for \HITrees: first, for execution (\autoref{sec:eval}) and second, for reasoning (\autoref{sec:exec}).

\begin{figure}
  \centering
\begin{minted}{lean4}
class MHandler (E E' : Effect) (ρ m) [Monad m] where
  handle : E.I → (k : E.O → HITree E' ρ) → m (HITree E' ρ)

instance sumMH {E₁ E₂ E' ρ m} [Monad m]
      [mh₁ : MHandler E₁ E' ρ m] [mh₂ : MHandler E₂ E' ρ m]
    : MHandler (E₁ ⊕ₑ E₂) E' ρ m where
  handle i k := match i with
    | .inl i => MHandler.handle i (k ∘ Sum.inl)
    | .inr i => MHandler.handle i (k ∘ Sum.inr)

instance unfoldMH {ρ m E EPre E'} [Monad m]
      [UnfoldEffect E EPre] [MHandler EPre E' ρ m]
      : MHandler E E' ρ m where
  handle i k := mh.handle (UnfoldEffect.invI i)
    (λ x => k (Eq.mp UnfoldEffect.out_eq x))

partial def eval {E ρ m} [Monad m] [MHandler E E ρ m]
    [Inhabited ρ] : HITree E ρ → m ρ
  | .pure r => return r
  | .impure i k => do let r ← MHandler.handle i k; eval r
  | .unreachable => panic! "Unreachable computation"
\end{minted}
  \caption{The \iLean{eval} interpretation}
  \label{fig:interp:eval}
\end{figure}

\subsection{Execution using Monads}
\label{sec:eval}
The first interpretation enables the execution of \HITrees by interpreting them into (executable) monads, provided by Lean's \iLean{Monad} typeclass.
Central to this interpretation is the \iLean{MHandler} typeclass (\autoref{fig:interp:eval}), which enables modular interpretation of different effects. An instance of \iLean{MHandler E E' ρ m} shows how to interpret any impure computation with input \iLean{i : E.I} and continuation \iLean{k} of type \iLean{E.O → HITree E' ρ} into the monad \iLean{m}.
As a concrete example, consider the \iLean{MHandler} instance for \iLean{StateE} from \autoref{sec:semantics-example}:
\begin{minted}{lean4}
instance {α E' ρ m} [Monad m] [MonadState α m]
  : MHandler (StateE α) E' ρ m where
  handle i k := match i with
    | .get => do return k (StateE.O.get (← MonadState.get))
    | .set v => do MonadState.set v; return k (StateE.O.set)
\end{minted}
This handler interprets \iLean{StateE α} into an arbitrary monad \iLean{m} that provides a state of type \iLean{α}. (This requirement is encoded using \iLean{[MonadState α m]}.)
The handler checks whether the input \iLean{i} is a \iLean{get} or \iLean{set} and forwards it to the \iLean{get} resp. \iLean{set} of the underlying monad.

Such individual handlers can be lifted to a composite effect like \iLean{ExE} using the sum handler \iLean{sumMH} (\autoref{fig:interp:eval}). This handler lifts \iLean{MHandler}s for \iLean{E₁} and \iLean{E₂} to an \iLean{MHandler} for \iLean{E₁ ⊕ₑ E₂} by matching on the sum in the input and then invoking the respective handler. For effect fixpoints, the \iLean{unfoldMH} handler lifts the \iLean{MHandler} for \iLean{EPre} to the \iLean{MHandler} for \iLean{E} given \iLean{UnfoldEffect E EPre}.

These \iLean{MHandler}s are used by the \iLean{eval} function to interpret a \HITree into a monad \iLean{m}.
\iLean{eval} recursively pattern matches on the \HITree:
for \iLean{.pure r}, \iLean{eval} returns \iLean{r};
for \iLean{.impure i k}, \iLean{eval} invokes the \iLean{MHandler} to interpret the effect and continues with the resulting \HITree; %
for \iLean{.unreachable}, \iLean{eval} terminates the interpretation by panicking.

\paragraph{Simple \iLean{MHandler}}
The \iLean{handle} function of \iLean{MHandler} is very powerful:
It takes as argument the full continuation \iLean{k} and can return an arbitrary \HITree, not just a version of \iLean{k}.
We will see in \autoref{sec:effects} how this allows interpreting effects that alter the control flow like concurrency.
However, for many effects a simpler variant suffices:
\begin{minted}{lean4}
class SMHandler (E m) [Monad m] where
  handle : E.I → m E.O
instance [Monad m] [SMHandler E m] : MHandler E E' ρ m where
  handle i k := do return k (← SMHandler.handle i)
\end{minted}
\iLean{SMHandler.handle} interprets \iLean{E.I} directly to \iLean{E.O} using \iLean{m}.
\iLean{SMHandler} makes it simpler to define \iLean{MHandler} for effects like \iLean{StateE}.

\subsection{Reasoning using State Machine Semantics}
\label{sec:exec}
\begin{figure}
  \centering
\begin{minted}{lean4}
class EHandler (E E' : Effect) (σ ρ) where
  handle : (i : E.I) → (s : σ) → (k : E.O→HITree E' ρ) →
           (p : HITree E' ρ → σ → Prop) → Prop
  handle_mono : ∀ {i s k p q},
    (∀ o s', p o s' → q o s') →
    handle i s k p → handle i s k q

inductive execPre exec : HITree E ρ → σ →
      (HITree E ρ → σ → Prop) → Prop where
  | halt t s p : p t s → execPre exec t s p
  | step i k s :
      EHandler.handle i s k (λ t' s' => exec t' s' p) →
      execPre exec (HITree.impure i k) s p

def exec (t : HITree E ρ) (s : σ)
    (p : HITree E ρ → σ → Prop) : Prop :=
  execPre exec t s p
coinductive_fixpoint monotonicity fun f f' himp => by ...
\end{minted}

  \caption{The \iLean{exec} interpretation}
  \label{fig:interp:exec}
\end{figure}
This section describes how we adapt the technique from \citet{ProgramLogicsALaCarte} to interpret \HITrees as state machines.%
\footnote{\citet{ProgramLogicsALaCarte} only describe a version of the technique that does not provide access to the continuation in the paper, but provide the full version in their Rocq development. We follow their Rocq development.}
These state machines can be seen as a big-step semantics for \HITrees and can be used to reason about all possible executions of a \HITree.

We follow the same recipe as \autoref{sec:eval}: We first define a notion of handlers \iLean{EHandler} that describes how to interpret an individual effect into a small-step relation.
Then, we define a function \iLean{exec} that lifts these handlers to a big-step relation on \HITrees. (We also write $(t, s) \Downarrow T$ for \iLean{exec t s T}.)
The definitions are shown in \autoref{fig:interp:exec}.

The definition of \iLean{EHandler} follows the structure of \iLean{MHandler}: \iLean{EHandler} is parameterized by (1) the effect \iLean{E} whose semantics is  being defined, (2) the effect \iLean{E′} and return type \iLean{ρ} of the continuation, and (3) the state \iLean{σ} of the \iLean{EHandler}.
The \iLean{handle} predicate gives the small-step relation for the input \iLean{i : E.I}, state \iLean{s : σ}, and continuation \iLean{k}. The final states of the step are given by the set \iLean{p}, following Omnisemantics~\cite{Omnisemantics}).
\iLean{handle_mono} states that
\iLean{handle} is monotonic in \iLean{p}.

Consider the \iLean{EHandler} for \iLean{DemonicE} from \autoref{sec:keyideas} as an example:
\begin{minted}{lean4}
instance {α E ρ} : EHandler (DemonicE α) E Unit ρ where
  handle i s k p := ∃ x : α, i.1 x ∧
                       p (k (DemonicE.O.choose x)) ()
  handle_mono := by grind
\end{minted}
To construct an execution of a demonic choice we must show that there \emph{exists} a choice \iLean{x} satisfying the predicate \iLean{i.1} passed as the input to the choice (see \autoref{sec:semantics-example}).
The final state is the continuation \iLean{k} applied to \iLean{x} (with unit for the state).

To define \iLean{exec}, we first introduce the predicate \iLean{execPre} that models one step of the execution.
There are two cases:
Either the execution halts using \iLean{execPre.halt} by showing that \iLean{p} contains the current \HITree \iLean{t} and state \iLean{s}.
Or if \iLean{t} is \iLean{.impure i k}, \iLean{execPre.step} executes a small step using \iLean{EHandler.handle} and then continues the execution for the resulting \HITree and state.
\iLean{exec} is then defined as the co-inductive fixpoint of \iLean{execPre}.
The use of co-induction means that \iLean{exec} can represent infinite executions.
Using the \iLean{coinductive_fixpoint} command requires proving that \iLean{exec} is monotonic, which follows from \iLean{handle_mono}.

Following \citet{ProgramLogicsALaCarte}, we obtain rules for reasoning about \iLean{exec}, in particular for co-induction, bind, and lifting \iLean{EHandler} through effect composition.

\section{Effects}
\label{sec:effects}

\begin{figure}
  \centering
\begin{subfigure}{\columnwidth}
\begin{minted}{lean4}
private structure FId where id : Nat
effect RecE (α β : Type) (E : Effect) where
  | fix : (_ : FId → α → HITree E β) (_ : α) → β
  | call : (_ : FId) (_ : α) → β

instance {ρ E α β} : EHandler (RecE α β E) E
    (List (α → HITree E β)) ρ where
  handle i s k p := match i with
    | RecE.I.fix f a => let f' := f ⟨s.length⟩;
       p (do let r ← f' a; k (RecE.O.fix r)) (s ++ [f'])
    | RecE.I.call fid a => match s[fid.id]? with
       | some f => p (do let r ← f a; k (RecE.O.call r)) s
       | none   => True
  handle_mono := by grind

instance {m ρ E α β} [RecE α β E -< E] [Monad m]
   [MonadStateOf (List (α → HITree E β)) m] :
   MHandler (RecE α β E) E ρ m where
  handle i k := do
    let s ← get; match i with
    | RecE.I.fix f a =>
      let f' := f ⟨s.length⟩; set (s ++ [f'])
      return (do let r ← f' a; k (RecE.O.fix r))
    | RecE.I.call fid a => match s[fid.id]? with
      | some f => return (do let r ← f a; k (RecE.O.call r))
      | none => panic! "Invalid FId"
\end{minted}
  \caption{The recursion effect.}
  \label{fig:effects:rec}
\end{subfigure}

\begin{subfigure}{\columnwidth}
\vspace{\subfigurespace}
\begin{minted}{lean4}
private structure KId where id : Nat
effect CallccE (α : Type) (E : Effect) where
  | callcc (f : KId → HITree E Empty) : α
  | throw (v : α) (k : KId) : Empty

instance {α E ρ} : EHandler (CallccE α E) E
    (List (α → HITree E ρ)) ρ where
  handle i s k p := match i with
    | .callcc f => let id := ⟨s.length⟩
      p ((f id).bind Empty.elim) (s ++ [λ v => k (.callcc v)])
    | .throw v k => match s[k.id]? with
      | some k => p (k v) s
      | none => True
-- MHandler is analogous
\end{minted}
  \caption{The call/cc effect.}
  \label{fig:effects:callcc}
\end{subfigure}

  \caption{Effects and their interpretations.}
  \label{fig:effects}
\end{figure}

\begin{figure}

\begin{minted}{lean4}
abbrev TId := Nat
private inductive Thread α E ρ where
  | yielded (t : HITree E ρ) | completed (val : α)
  | blocked (b₁ : TId) (b₂ : TId) (k : ConcE.O α → HITree E ρ)
abbrev ThreadPool α E ρ := List (Thread α E ρ)

private def Thread.continue? (th : Thread α E ρ)
    (tp : ThreadPool α E ρ) : Option (HITree E ρ) :=
/- check if th is yielded or its blocking threads completed -/

structure HandlerState α E ρ where
  curr : TId
  tp : ThreadPool α E ρ

/- EHandler -/
def choose_thread (tp : ThreadPool α E ρ) p :=
  ∃ nextId nextThread t, tp[nextId]? = some nextThread ∧
    nextThread.continue? tp = some t ∧ p t ⟨curr, tp⟩

instance : EHandler (ConcE α E) E (HandlerState α E ρ) ρ where
  handle i s k p :=
    match i with
    | .par t₁ t₂ => let ⟨curr, tp⟩ := s
      let t' := Thread.blocked tp.length tp.length.succ k
      let tp' := (tp.set curr t') ++ [.yielded t₁, .yielded t₂]
      choose_thread tp' p
    | .yield => let ⟨curr, tp⟩ := s
      let tp' := tp.set curr (.yielded (k .yield))
      choose_thread tp' p
    | .kill v => let ⟨curr, tp⟩ := s
      let tp' := tp.set curr (.completed v)
      choose_thread tp' p
  handle_mono := by ...

/- MHandler -/
def schedule (s : HandlerState α E ρ) :
  HandlerState α E ρ × HITree E ρ :=
/- schedule next unblocked thread in round-robin -/

instance [Monad m] [MonadStateOf (HandlerState α E ρ) m]
    [Inhabited α] : MHandler (ConcE α E) E ρ m where
  handle i k := match i with
    | .par t₁ t₂ => do
      let ⟨curr, tp⟩ ← get
      let t' := Thread.blocked tp.length tp.length.succ k
      let tp := (tp.set curr t') ++ [.yielded t₁, .yielded t₂]
      let ⟨s', t'⟩ := schedule {curr, tp}; set s'; return t'
    | .yield => do
      let ⟨curr, tp⟩ ← get
      let tp := tp.set curr (.yielded (k .yield))
      let ⟨s', t'⟩ := schedule {curr, tp}; set s'; return t'
    | .kill v => do
      let ⟨curr, tp⟩ ← get
      let tp := tp.set curr (.completed v)
      let ⟨s', t'⟩ := schedule {curr, tp}; set s'; return t'
\end{minted}
  \caption{Interpretation of the concurrency effect.}
  \label{fig:effects:conc}
\end{figure}

\autoref{sec:keyideas} showed the basics of defining and using effects. This section comes back to the most interesting higher-order effects and their interpretations.
In particular, we discuss \iLean{RecE} and \iLean{ConcE} from \autoref{sec:semantics-example} and introduce a call/cc effect.

\paragraph{Recursion}
\autoref{fig:effects:rec} shows the recursion effect \iLean{RecE}.
Recall from \autoref{sec:semantics-example} that \iLean{RecE} provides two operations:
\iLean{fix} to create a fixpoint and \iLean{call} to invoke a fixpoint.
Fixpoints are represented using the \iLean{FId} type, which serve as abstract identifiers (implemented by natural numbers).

Both the \iLean{EHandler} and \iLean{MHandler} for \iLean{RecE} follow the same strategy to implement the effect:
They store a list of \HITree functions representing the fixpoints, with the index in the list representing the id of the fixpoint. Invoking \iLean{fix} with a function \iLean{f} first provides the id of the new fixpoint to \iLean{f}.
Then it appends the resulting function \iLean{f'} to the list of fixpoints and continues with the \HITree executing \iLean{f'} and the \iLean{k}.
The handler for \iLean{call} looks up the corresponding fixpoint for an identifier (terminating the execution if it does not exist) and executes the fixpoint.
Together, these encode the desired behavior for recursion where one can create new recursive functions via \iLean{fix} that can invoke themselves using \iLean{call}.

\paragraph{Call/CC}
Control effects like call/cc are a notoriously subtle feature in higher-order languages~\cite{HigherOrderStateReasoning}.
To show that \HITrees can support control effects, we implement the \iLean{CallccE} effect in \autoref{fig:effects:callcc}.
\iLean{CallccE} provides two operations: \iLean{callcc} captures the current continuation and passes it as an abstract identifier \iLean{KId} to the function \iLean{f}. \iLean{throw} invokes the continuation \iLean{k} with value \iLean{v}.
(Note that we use defunctionalization via \iLean{KId} to deal with higher-order \HITrees, here the continuation, in a negative position.)
The interpretation of \iLean{CallccE} is straightforward since both \iLean{EHandler} and \iLean{MHandler} provide the \iLean{handle} function with the current continuation. The handlers store the continuations in a list in the state and invoke them on \iLean{throw}.

\paragraph{Concurrency}
The interpretations of the \iLean{ConcE} effect from \autoref{sec:semantics-example} are shown in \autoref{fig:effects:conc}.
First, we define the \iLean{ThreadPool} data structure used by both interpretations. A \iLean{ThreadPool} is a list of \iLean{Thread}s where each thread can either be \iLean{yielded} with body \iLean{t}, \iLean{completed} with resulting value \iLean{val}, or blocked on a parallel composition. A blocked thread tracks the thread identifiers \iLean{b₁} and \iLean{b₂} of the two threads of the parallel composition and the continuation \iLean{k} with the computation after the parallel composition.
The \iLean{Thread.continue?} function checks if the thread \iLean{th} can continue by checking that it is either yielded or it is blocked and the two blocking threads have completed.
The state \iLean{HandlerState} combines a \iLean{ThreadPool} with the id of the currently executing thread.

The \iLean{EHandler} picks the next thread using the \iLean{choose_thread} function. This function uses an existential quantifier to model the non-deterministic scheduling usually used by operational semantics.
The \iLean{EHandler} handles the three operations of \iLean{ConcE}: \iLean{par t₁ t₂} marks the current thread as blocked and adds \iLean{t₁} and \iLean{t₂} as new threads that can be executed, before choosing the next thread.
\iLean{yield} updates the current thread in the thread pool to \iLean{k} and then chooses a new thread.
\iLean{kill} sets the current thread to \iLean{completed}. The \iLean{choose_thread} will not pick this thread again since it cannot continue.
Overall, this handler allows us to reason about all possible concurrent executions using a non-deterministic thread pool semantics.

For the \iLean{MHandler}, the \iLean{schedule} function implements a round-robin scheduler.
The \iLean{MHandler} follows \iLean{EHandler}, except that it uses \iLean{schedule} as a concrete implementation of a scheduler.
This allows us to evaluate and test concurrent \HITrees.
In the future, it would be interesting to explore different scheduling strategies.%

\section{Case study: \SimpLang}
\label{sec:case-study}

\begin{figure}
  \centering
      \begin{align*}
        v \in \valset& \Coloneqq \grayout{z \mid \ell \mid (v_1, v_2)} \mid k \mid \grayout{\Lam x. e} \quad (\grayout{z \in \mathbb Z, \ell,} k \in \mathbb N)
        \\
        e \in \exprset& \Coloneqq \grayout{v \mid x \mid e_1 \langplus e_2} \mid e_1 \langeq e_2 \mid \grayout{e.1 \mid e.2 \mid  \deref{e} \mid e_1 \gets e_2 \mid} \\
        & \grayout{\alloc(e) \mid \parcomp{e_1}{e_2} \mid e_1(e_2)} \mid \callcc{e} \mid \throw{e} \mid \\
        & \asrt{e} \mid \faa{e} \mid \If e_1 then e_2 \Else e_3
    \end{align*}
    \[
      \Let x = e₁ in e₂ \eqdef (\Lam x. e₂)(e₁) \quad e₁; e₂ \eqdef \Let \_ = e₁ in e₂
    \]
  \caption{\SimpLang, parts shared with \ExampleLang in \grayout{gray}.}
  \label{fig:simplang}
\end{figure}

To evaluate \HITrees and their interpretations, we use them to define the semantics of the lambda calculus \SimpLang combining concurrency and call/cc.%
\footnote{Our implementation of \SimpLang is inspired by SimpLang~\cite{SimpLang}.}
The syntax of \SimpLang is shown in \autoref{fig:simplang}.
\SimpLang extends \ExampleLang from \autoref{sec:keyideas} with call/cc by adding the $\callcc{e}$ and $\throw{e}$ expressions. Captured continuations are represented as the values $k$.
\SimpLang adds an equality operation $e_1 \langeq e_2$, an $\asrt{e}$ expression that checks that $e$ evaluates to a non-zero integers, an atomic fetch-and-add $\faa{e}$, and if-expressions. Let-bindings and sequencing are defined as standard.

\paragraph{Denotational semantics}
We define the denotational semantics of \SimpLang based on the following effect:
\begin{minted}{lean4}
fix_effect E := StateE (ExtTreeMap Loc Val) ⊕ₑ CallccE Val E
   ⊕ₑ RecE Exp Val E ⊕ₑ ConcE Val E ⊕ₑ FailE ⊕ₑ DemonicE Loc
\end{minted}
The effect \iLean{E} provides the effects necessary to model \SimpLang:
\iLean{StateE} to model the heap,
\iLean{CallccE} for $\callcc{e}$,
\iLean{RecE} to represent recursive functions,
\iLean{ConCE} for concurrency,
\iLean{FailE} to model failing executions,
and \iLean{DemonicE} for non-deterministic allocation.
Using this effect, we define a recursive \iLean{denote} function similar to the \iLean{denote} function from \autoref{sec:keyideas}.
$\callcc{e}$ and $\throw{e}$ dispatch to the corresponding operations of the \iLean{CallccE} effect.
The placement of \iLean{yield}s follows \citet{ProgramLogicsALaCarte}.

\begin{figure}
  \centering
\begin{minted}{lean4}
def awk := [λ|
  let x := ref #0 in
  λ f, (x ← #0; f #(); x ← #1; f #(); assert (!x = #1))
]
def C_callcc := [λ|
  let g := $awk in let b := ref #0 in
  let f := (λ _, if !b then -- ! is a load, not negation
    call/cc (λ k, g (λ _, throw #() to k)) else b ← #1) in
  g f
]
#eval eval C_callcc -- fails with assertion failure
theorem callcc p : exec C_callcc.denote default p := by ...

def C_conc := [λ|
  let g := $awk in let f := λ _, #() in g f || g f
]
#eval eval C_conc -- succeeds
theorem conc p : exec C_conc.denote default p := by ...
\end{minted}
  \caption{Very awkward example.}
  \label{fig:awk}
\end{figure}

\paragraph{Very awkward example}
From this denotational semantics
we automatically obtain the evaluation and state machine interpretations described in \autoref{sec:interpretations} by composing the handlers for the individual effects.
Our Lean development uses a few small examples to test these interpretations. Here we focus on the most interesting of these examples: the very awkward example~\cite{HigherOrderStateReasoning} shown in \autoref{fig:awk}. (We use Lean's extensible syntax mechanisms~\cite{LeanMacros} to embed \SimpLang programs inside \iLean{[λ| ...]}. \iLean{$awk} %
embeds the definition of \iLean{awk} inside \iLean{C_callcc} and \iLean{C_conc}.)

The very awkward example tests whether a language enforces well-bracketed control-flow. Concretely, for a language with well-bracketed control flow, there is no context that makes the assertion in \iLean{awk} fail~\cite{HigherOrderStateReasoning}.
However, \SimpLang provides two mechanisms that break well-bracketed control flow: call/cc and concurrency.
We provide two contexts, \iLean{C_callcc} and \iLean{C_conc}, that use these features respectively to violate the assertion in \iLean{awk}.

\iLean{C_callcc} uses call/cc to capture the assert in the continuation the second time \iLean{f} is invoked, and then jump to it after \iLean{x} has been set to 0 in a second invocation of \iLean{g}.
Luckily, we do not need to understand these details to see that \iLean{C_callcc} violates the assertion---we can just use the \iLean{eval} interpretation to execute the \HITree.
This evaluation indeed triggers the assertion failure.
We also use \iLean{exec} to prove that there is an execution of \iLean{C_callcc} where the assertion fails.
Concretely, we show that we can prove \iLean{exec C_callcc.denote default p} where \iLean{default} is the initial state of \SimpLang and \iLean{p} is an arbitrary postcondition.
The fact that we can prove this for an arbitrary \iLean{p} means that the behavior is unconstrained, \ie that the program exhibits the undefined behavior caused by a failed assertion.

For \iLean{C_conc}, the evaluation using \iLean{eval} does not trigger the failed assertion since the round-robin scheduler does not pick the right interleaving for this. We show using \iLean{exec} that there in fact exists a schedule that violates the assertion (the schedule where the \iLean{x ← 0} of the one thread is executed right before the \iLean{assert (!x = 1)} of the other thread).

\section{Related Work}
\label{sec:related-work}

This section first compares the design and trade-offs of \HITrees to ITrees~\cite{InteractionTrees} before discussing related work more broadly.

\paragraph{\HITrees vs. ITrees}
We compare \HITrees and ITrees on three different axes: the notion of effects, the use of induction vs. co-induction, and the reasoning styles.

The key feature of \HITrees is their support for higher-order effects like \iLean{ConcE} (\autoref{sec:keyideas}) that cannot be directly represented in ITrees.
However, \HITrees have to pay a price: For \HITree effects (unlike ITree effects) the type of outputs cannot depend on the input (statically).
\autoref{sec:keyideas} demonstrated how to encode such dependencies dynamically using \iLean{.unreachable}.
But this requires a decidable dynamic cast and there are effects, like try 1 of \iLean{DemonicE} in \autoref{sec:semantics-example} that cannot be encoded with a dynamic cast (in this case, since equality on types is not decidable).%
\footnote{One can use excluded middle to avoid the decidability requirement, but this makes the \HITree noncomputable.}
\autoref{sec:semantics-example} shows how to obtain a version of \iLean{DemonicE} suitable for defining the semantics of allocation.
\iLean{DemonicE} (and its dual \iLean{AngelicE}) are the only effects where we encountered issues with providing decidable dynamic casts, but there might be other problematic dependent effects that we have not considered.

Another difference of \HITrees compared to ITrees is that \HITrees are defined as an inductive type instead of a co-inductive type. This is mostly due to the fact that Lean does not have native support for co-inductive types. (We do not foresee a technical problem with building a co-inductive version of \HITrees in Rocq.) Using an inductive type makes the meta-theory of \HITrees simpler since it avoids the need for the Tau constructor of ITrees. This means that \HITrees can use the standard notion of equality instead of the equality up to (finitely many) Tau used by ITrees.
On the flip side, \HITrees cannot natively represent infinite computations. Instead, \HITrees rely on effects like \iLean{rec} from \autoref{sec:semantics-example} to encode loops.

An important difference between ITrees and \HITrees is how one reasons about them. \citet{InteractionTrees} provide a rich theory of \emph{equational} reasoning principles for ITrees and use them to reason about the step-wise interpretation of effects into other ITrees.
In contrast, such step-wise interpretation is more challenging for \HITrees since they cannot natively represent infinite computations (see above) and thus \HITrees cannot represent the interpretation of effects like \iLean{RecE}. As a consequence, the equational theory of \HITrees cannot be used to reason about the interpretation of effects. In particular, we do \emph{not} have $\mathit{rec}~f = f~(\mathit{rec}~f)$. ($\mathit{rec}$ only becomes meaningful when paired with an interpretation (\autoref{sec:interpretations}, \autoref{sec:effects}).)
Instead of relying on equational reasoning and step-wise interpretation,  we follow \citet{ProgramLogicsALaCarte} and \citet{Osiris} and directly interpret \HITrees into a rich domain---Lean's monads or a state machine semantics. \autoref{sec:interpretations} and \autoref{sec:effects} show how this allows us to model even complex effects like demonic non-determinism, call/cc, or concurrency.

\paragraph{Related work}
Defining denotational semantics using free monads goes back to the Datatypes \`a la carte paper~\cite{DatatypesALaCarte} that popularized free monads in Haskell. Subsequent work on the free-er monad~\cite{FreerMonad1, FreerMonad2, FreerMonad} refined the definition of free monads by exposing the continuation in the \iLean{impure} constructor.
This definition was adapted by \citet{FreeSpec} and \citet{InteractionTrees} into the FreeSpec resp. interaction trees libraries inside the Rocq proof assistant, with FreeSpec using an inductive definition focusing on modeling first-order, low-level devices, while interaction trees use a co-inductive definition with a rich equational theory. Subsequent work extended interaction trees to represent non-deterministic programs~\cite{ChoiceTrees}.
\citet{ITreeIsabelle} show how to implement a version of interaction trees in Isabelle/HOL.
To our knowledge, none of this work on free monads in proof assistants considered higher-order effects. It is unclear how interaction trees could support higher-order effects due to the negative occurrence of the input type in the definition of ITrees.

We are only aware of one version of interaction trees that support higher-order effects: guarded interaction trees~\cite{GITrees, GITreesContext}.
Guarded interaction trees leverage a guarded type theory to support higher-order inputs and outputs. They show that this makes GITrees a powerful denotational semantic that allows interoperation between different higher-order languages.
However, working inside the guarded type theory instead of the ambient type theory of the proof assistant makes GITrees more cumbersome to use (\eg by requiring reasoning about step-indexing).
Instead, we show how one can encode a higher-order version of interaction trees inside the ambient type theory of a proof assistant.
In the future, it would be interesting to investigate whether \HITrees can be used as a suitable denotational semantics that allows interoperation between different languages.

\citet{ProgramLogicsALaCarte} show how one can build a modular program logic based on ITrees. Their state machine interpretation of ITree is the bases the interpretation of \HITrees described in \autoref{sec:exec}. In the future, we are interested in investigating whether one can adapt their separation-logic-based program logic to \HITrees.

\begin{acks}                            %
  We thank Thibaut Perami, Amin Timany, and Lars Birkedal for the insightful discussions about this work.
\end{acks}

{
\interlinepenalty=10000
\bibliography{bib}
}

\end{document}